\documentclass{ws-procs9x6}

\newcommand{\apj}[3]{Astrophys.\ J.\ {\bf #1} (#2) #3}

\newcommand{\prd}[3]{Phys.\ Rev.\ {\bf D#1} (#2) #3}

\begin{document}

\title{Indirect searches for neutralino dark matter}

\author{Joakim Edsj\"o}

\address{Department of Physics, Stockholm University\\
AlbaNova, SE-106 91 Stockholm, Sweden\\
E-mail: edsjo@physto.se}


\maketitle

\abstracts{
There is mounting evidence for dark matter in the Universe and one of the favourite dark
matter candidates is the neutralino, which naturally appears as the
lightest supersymmetric particle (LSP) in many supersymmetric extensions of the standard model. The neutralino has the desired properties to be a good dark matter candidate and
we will here review the different indirect searches for neutralino dark matter and discuss the implications on these from recent direct searches.}

\section{Introduction and supersymmetric model}

Neutralinos arise naturally as good dark matter candidates in supersymmetric extensions of the standard model. We will here work in the Minimal Supersymmetric Standard Model (MSSM), with the usual low-energy parameters: $\mu$, $M_2$, $\tan \beta$, $m_A$, $m_0$,
$A_b$ and $A_t$.
See \cite{bg,coann} for more details.

We will use {\sffamily DarkSUSY}\cite{darksusy} to calculate the various rates in indirect and direct searches. The relic density of neutralinos is calculated including coannihilations between neutralinos and charginos\cite{coann,GondoloGelmini}.
We will here only include cosmologically interesting models, where the neutralinos can make up a major part of the dark matter in the Universe without overclosing it.

\section{Indirect searches and comparison with direct searches}

There are many different ways to search for neutralino dark matter (for a review see e.g.\ \cite{lars-review,jkg}). We will here focus on a few of the more promising where
experimental efforts are being done.

\subsection{Positrons}

\begin{figure}[t]
\centerline{\epsfxsize=0.49\textwidth\epsfbox{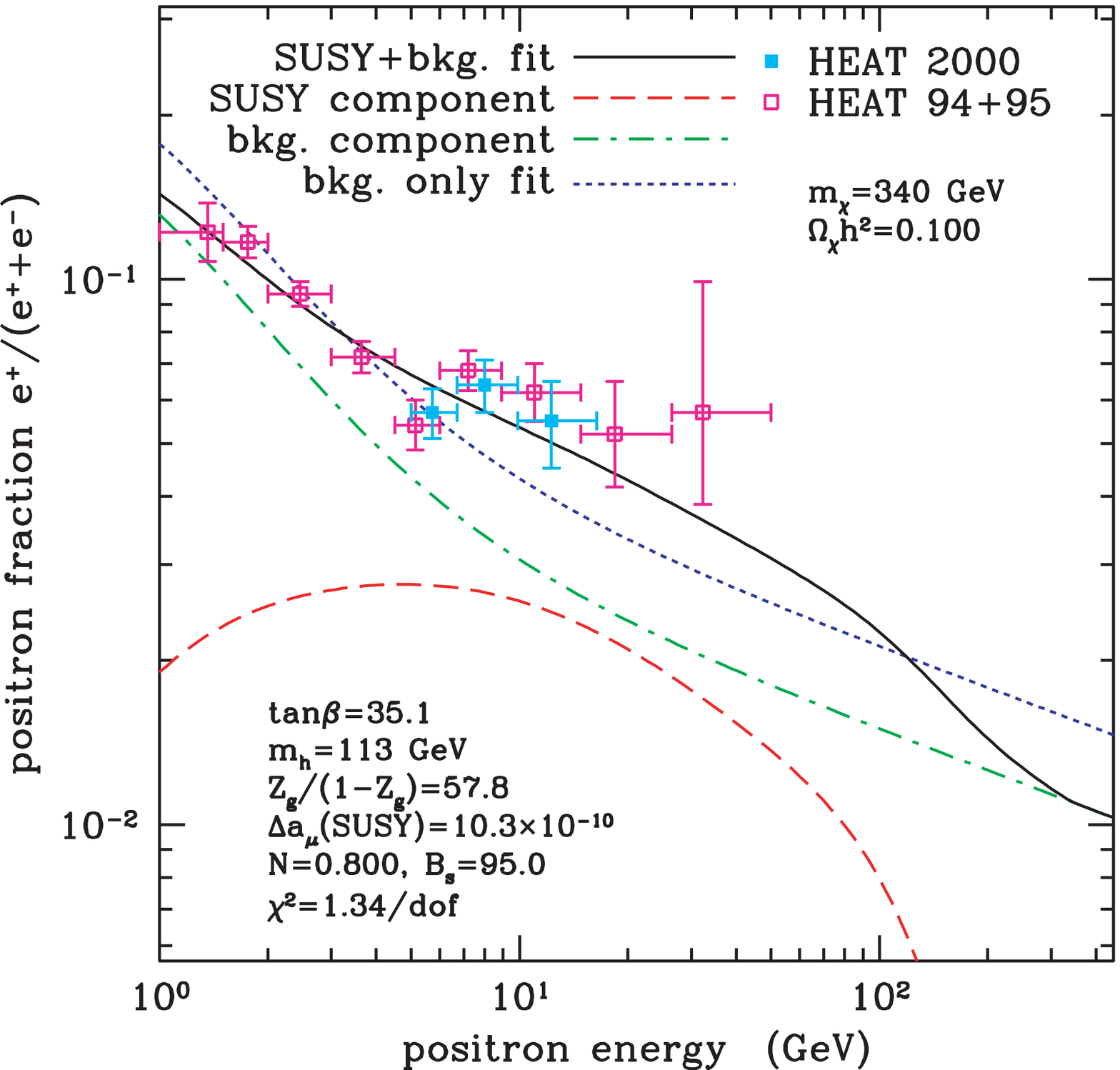}
\epsfxsize=0.49\textwidth\epsfbox{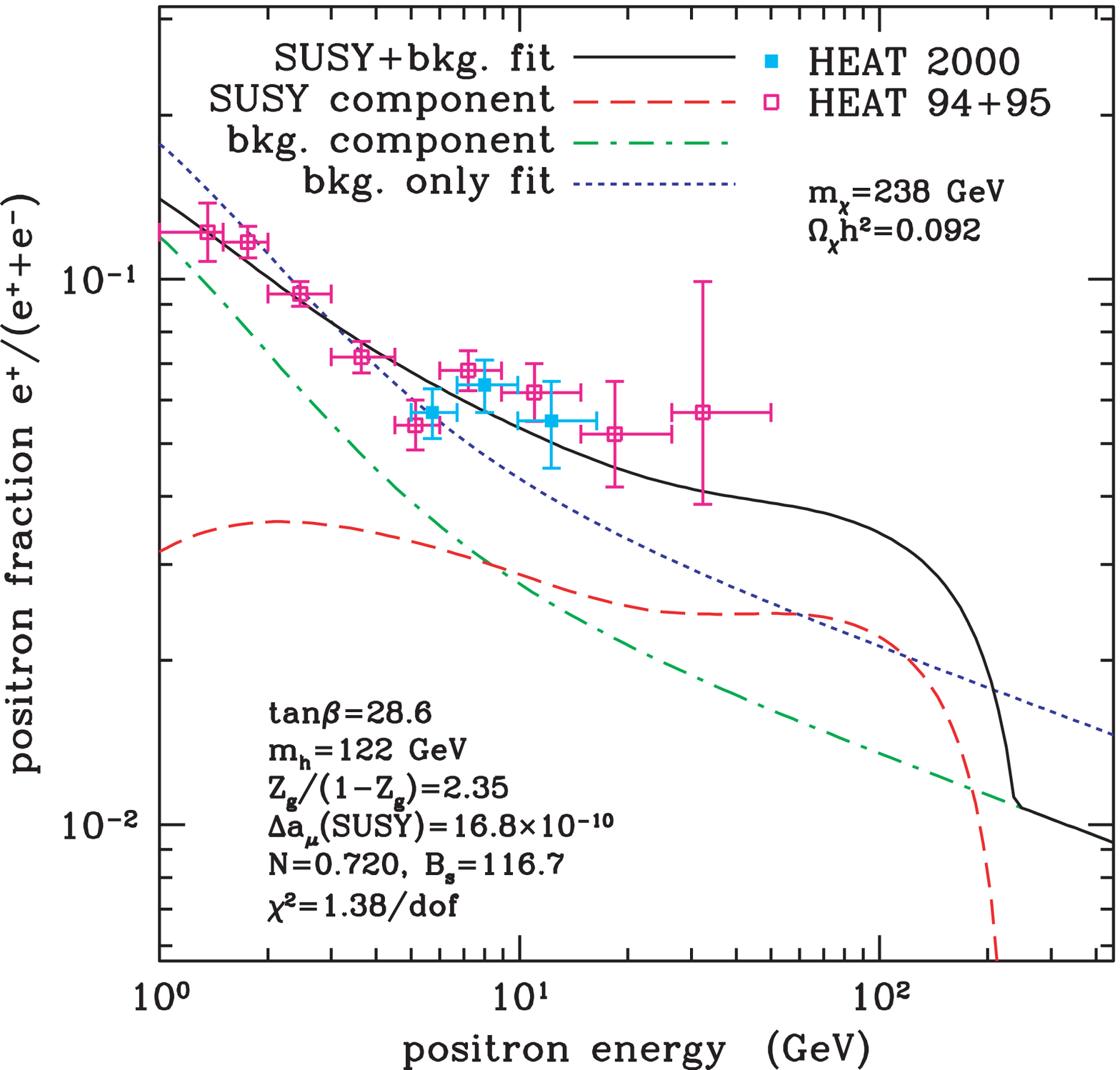}}   
\caption{The positron fraction as measured by HEAT 94+95\protect\cite{heatfrac}
and HEAT 2000\protect\cite{newdata}. Also illustrated is a background only fit and
a SUSY+background fit for two MSSM models with good fits. Both of these models
are gaugino dominated and the model in a) has positrons primarily from
hadronization, while the model in b) has hard positrons from direct gauge boson
decays.}
\label{fig:positrons}
\end{figure}

Neutralinos can annihilate in the galactic halo producing positrons\cite{positrons,positrons-other,kane-positrons}. 
The positron fraction has been measured by e.g.\ the HEAT Collaboration\cite{heatfrac,newdata} were an intriguing excess at $\sim8$ GeV has been seen. For a typical MSSM model, the predicted positron flux is too low to be seen, but if it is enhanced (by e.g.\ a clumpy halo) by a factor $\geq 30$ we can fit the observed flux. In Fig.~\ref{fig:positrons}, two models with good fits to the HEAT data are shown. We see that the fits are better than the background-only fits, but we do not reproduce the sharp bump at $\sim8$ GeV as indicated by the data. Not even a monochromatic source would produce such a sharp feature\cite{kane-positrons}. These fits do, however, produce features in the spectrum at higher energies that can be searches for with future detectors.

\subsection{Antiprotons}

\begin{figure}[t]
\centerline{\epsfxsize=0.49\textwidth\epsfbox{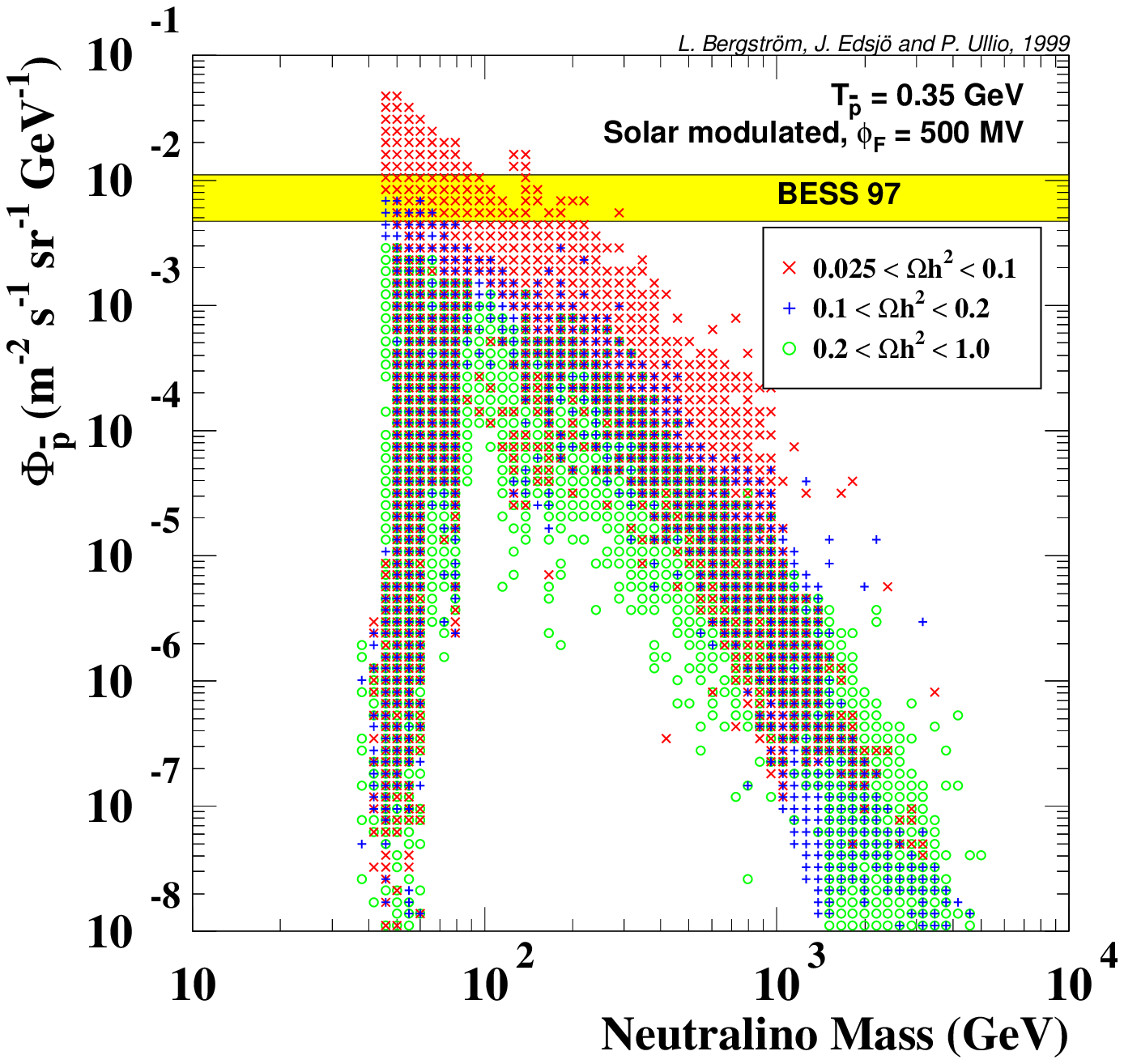}
\epsfxsize=0.49\textwidth\epsfbox{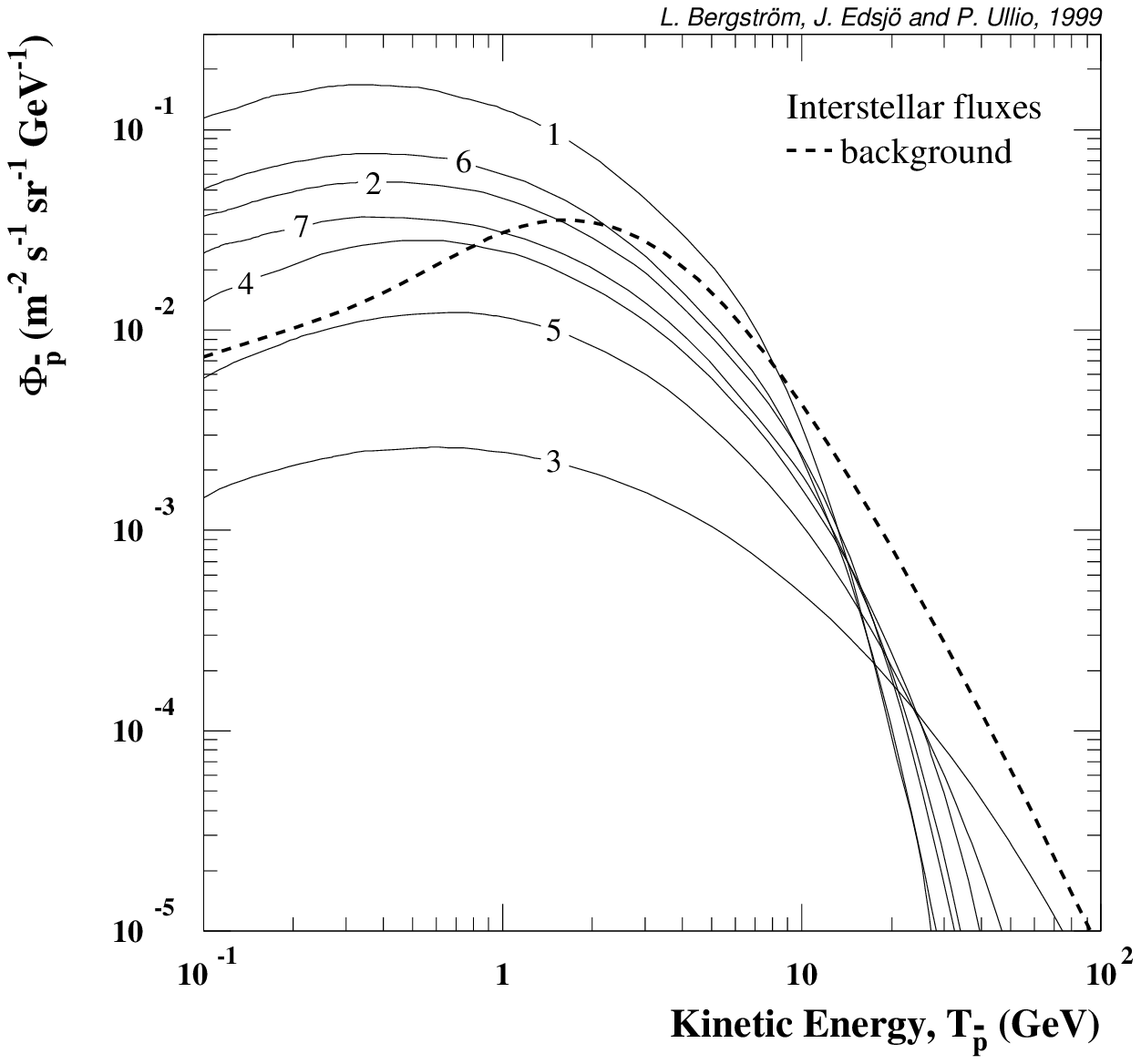}}   
\caption{The solar-modulated antiproton flux at 0.35 GeV kinetic energy compared with
the BESS 97 measurement\protect\cite{bess97}.}
\label{fig:antiprotons}
\end{figure}

Compared to positrons, it is easy to produce high fluxes of antiprotons from neutralino annihilation in the galactic halo. Unfortunately, they do not show an as clear signature as the positrons. In Fig.~\ref{fig:antiprotons} we show the antiproton flux (see \cite{pbar} for details) versus the neutralino mass and some example spectra compared
with the expected background. Even though we do not expect striking features in the antiproton spectra, they are easily overproduced in e.g.\ clumpy halo models and can put severe constraints on a given MSSM and astrophysical model. For the positron spectra in Fig.~\ref{fig:positrons} we have checked that those
models would not be ruled out by the antiproton measurements.

\subsection{Gamma rays}

\begin{figure}[t]
\centerline{\epsfxsize=0.49\textwidth\epsfbox{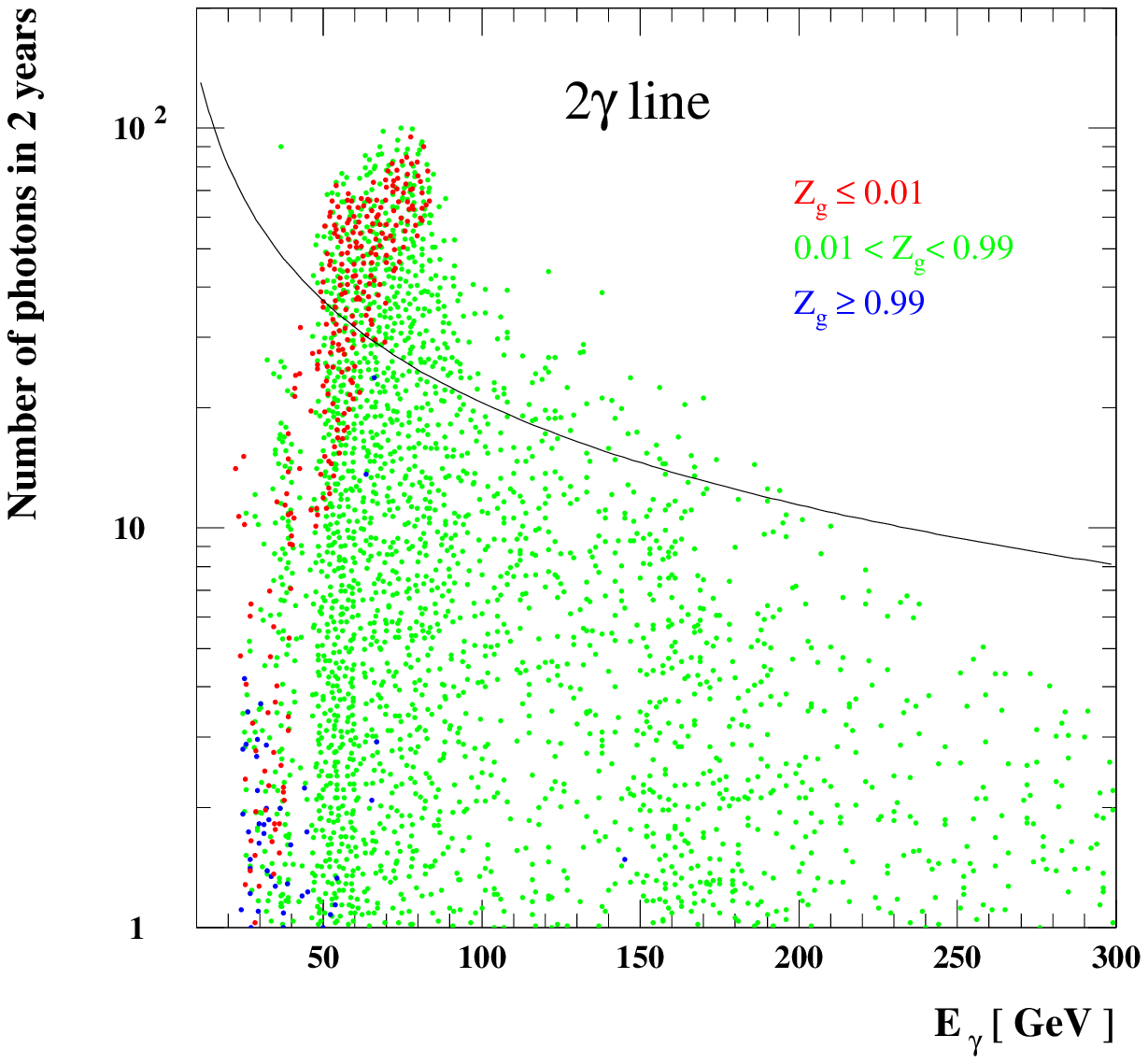}
\epsfxsize=0.49\textwidth\epsfbox{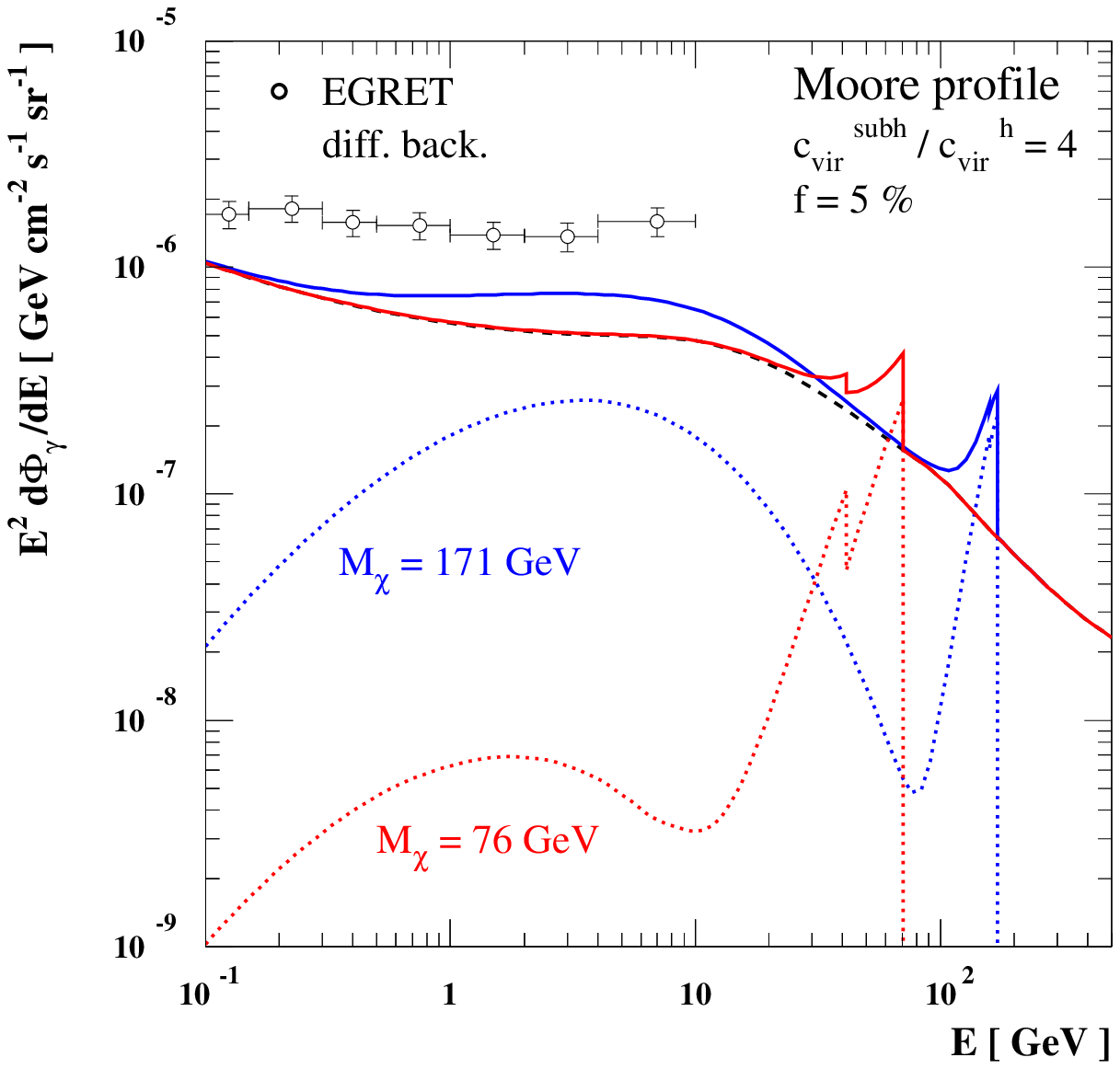}}   
\caption{a) Gamma rays from neutralino annihilation into $\gamma\gamma$ from the galactic center. Also shown are the expected sensitivity curves for GLAST. b) Diffuse gamma rays produced by annihilation in external halos (integrated over redshift). }
\label{fig:gammas}
\end{figure}

Gamma rays can be produced both monochromatically by annihilation (at the 1-loop level) to $\gamma \gamma$ and $Z \gamma$ and as a continuum from hadronization.
They are produced by annihilation in the halo and can produce high fluxes especially towards the galactic center. In Fig.~\ref{fig:gammas}a we show an example of expected monochromatic gamma ray fluxes towards the galactic center\cite{bub}. For some models, the fluxes would be detectable by e.g.\ GLAST.
However, these predicted fluxes are typically
very sensitive to the exact details of the halo model at the very center of the Milky Way. Another source of gamma rays would be to sum up the contribution from cosmological halos\cite{cosm-gamma}. In Fig.~\ref{fig:gammas}b we show two example spectra of the expected diffuse gamma ray fluxes as it could be seen by GLAST. The EGRET data points are shown for comparison\footnote{Note that the predictions are for the point source sensitivity of GLAST. With that of EGRET, the predictions would be a factor of two higher and would fit the data points well.}.
One can clearly see the monochromatic lines from annihilation into $\gamma\gamma$ and $Z\gamma$ where the triangular shape comes from the redshift.

\subsection{Neutrino-induced muons in neutrino telescopes}

\begin{figure}[t]
\centerline{\epsfxsize=0.49\textwidth\epsfbox{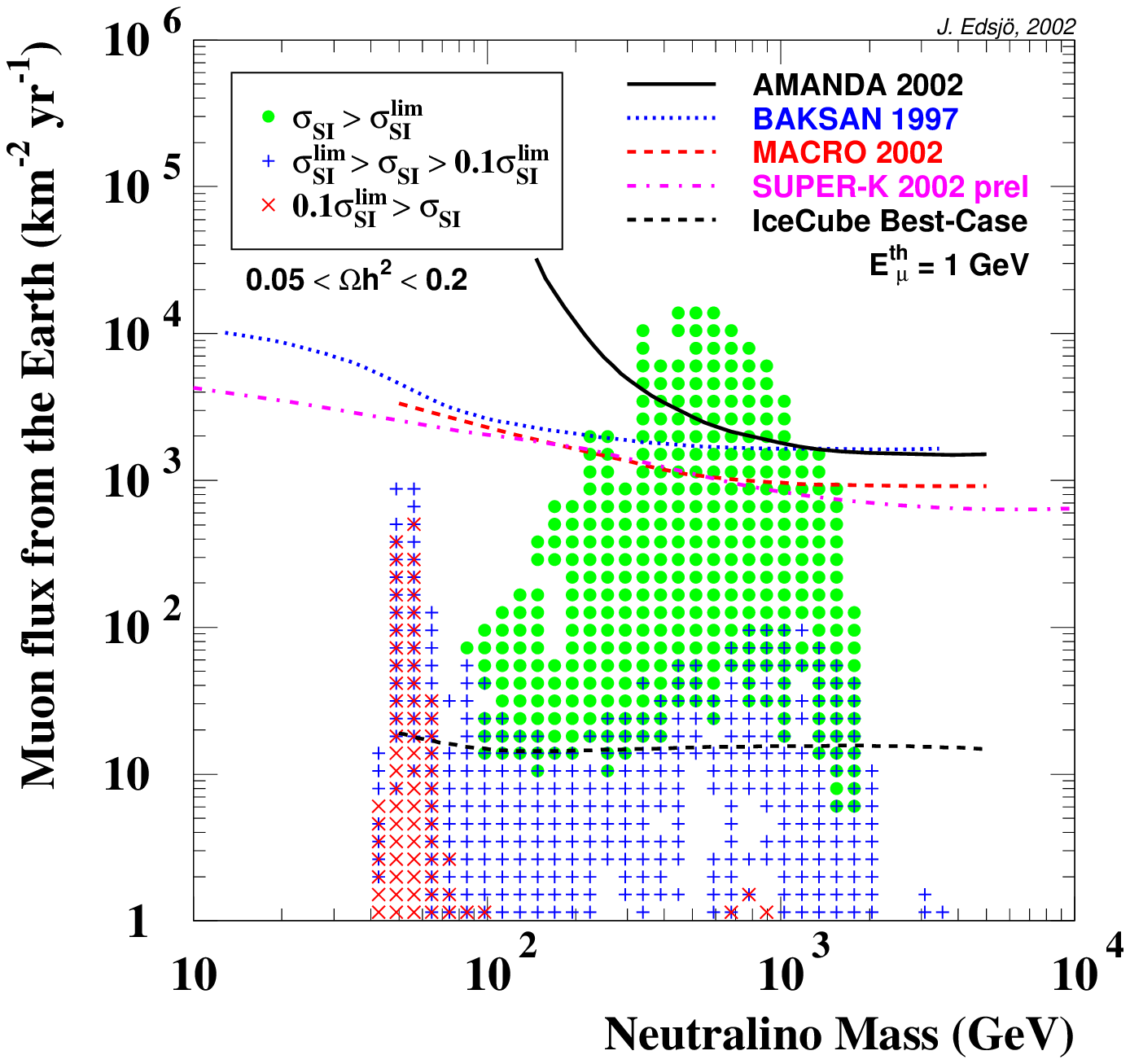}
\epsfxsize=0.49\textwidth\epsfbox{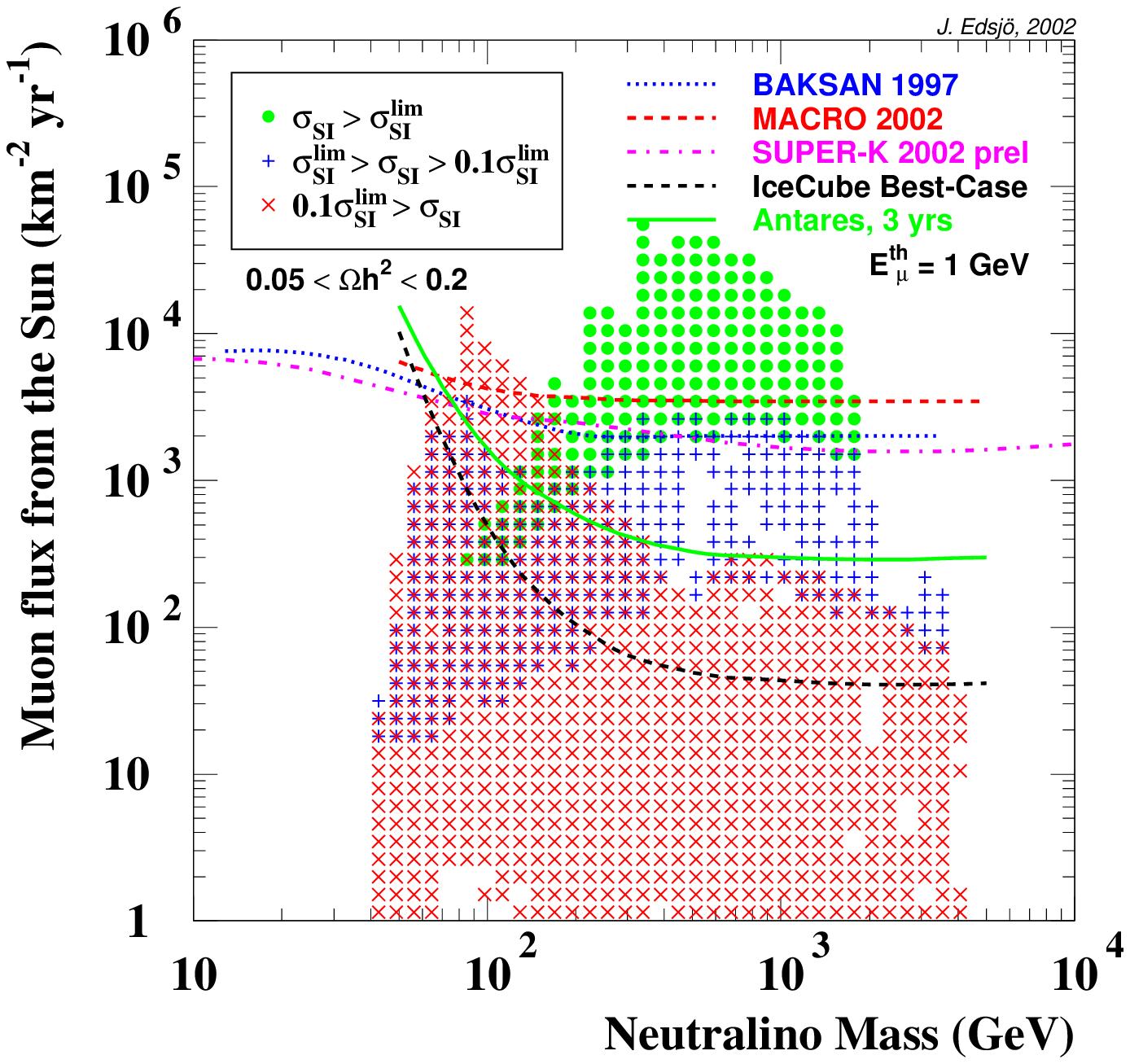}}   
\caption{Expected neutrino-induced muon fluxes from neutralino annihilation in a) the
Earth and b) the Sun. The current experimental limits are also shown together with the
expected future limits from Antares and IceCube. Models that are excluded by the current
Edelweiss limit are shown with green filled circles (see text for details).}
\label{fig:easu}
\end{figure}

Neutralinos can be gravitationally trapped by e.g.\ the Sun and the Earth, where they can annihilate and produce e.g.\ muon neutrinos. These can produce muons that neutrino-telescopes like Super-Kamiokande, Macro, Baksan, Amanda, Antares or IceCube could search for. In Fig.~\ref{fig:easu} we show the expected fluxes from neutralino annihilation in the Earth and the Sun\cite{neutel}. We also show the current limits from
Baksan\cite{baksan-lim}, Macro\cite{macro-lim}, Super-Kamiokande\cite{sk-lim} and Amanda\cite{ama-lim}. All limits have been converted to limits on the muon flux above 1 GeV (without an angular cut-off). We see that these neutrino telescopes have already started to explore the MSSM parameter space. We also indicate which models are excluded by the direct search by Edelweiss\cite{edelweiss}: green filled circles are excluded by Edelweiss, blue crosses would be excluded with a factor of ten increased sensitivity and red crosses would require an even higher sensitivity. We clearly see that for the Earth,
Edelweiss has already excluded those models that would produce the highest fluxes in neutrino telescopes, whereas for the Sun, the correlation is not as high. The reason for the strong correlation for the Earth is that both the signal in Edelweiss and the signal from the Earth depends strongly on the spin-independent scattering cross section. For the Sun, on the other hand, the signal depends strongly also on the spin-dependent scattering cross section, on which the limits are not as good from direct searches. One should keep in mind, however, that the correlation seen in Fig.~\ref{fig:easu} depends strongly on the assumed halo velocity profile (and for the Earth on the diffusion of neutralinos in the solar system\cite{gould-diff}). The capture by the Earth and the Sun is most efficient for low-velocity neutralinos whereas the direct detection rates are higher for high-velocity neutralinos. Hence, it could be possible to break this strong correlation in signal strengths with a different velocity profile. Fig.~\ref{fig:easu} is produced with the assumption of a standard gaussian halo profile.
Also indicated in the figure are the expected future limits from Antares\cite{antares} and IceCube.

\section{Conclusions}

The lightest supersymmetric particle is in many cases the lightest neutralino, which is an excellent dark matter candidate. We have shown that, within the MSSM, there are many
signals from neutralino dark matter that could be within reach of current or future detectors. The indication of an excess in the positron spectra as seen by HEAT\cite{heatfrac,newdata} is better explained by neutralinos annihilating in the galactic halo than by the conventional sources only, even though the fit is not perfect.
For many models, the antiproton fluxes are within reach of both current and future experiments. However, the antiproton fluxes do not show an as clear signature as the positron fluxes which make them hard to distinguish from the background. Gamma rays is another interesting signal, both from the galactic halo and from cosmological halos (making up the diffuse gamma ray background). Future gamma ray experiments like GLAST could search for these signals. Neutrino telescopes have already started to explore the MSSM parameter space and future detectors will improve the search potential, especially when looking towards the Sun, where the correlation with direct searches is not as strong.

\section*{Acknowledgments}
J.E.~thanks the Swedish Research Council for support.


\end{document}